# Protection of Hardware: Powering Systems (Power Converter, Normal Conducting, and Superconducting Magnets)

*H. Pfeffer, B. Flora, and D. Wolff*
US Particle Accelerator School, Batavia, IL, USA

**Abstract**
Along with the protection of magnets and power converters, we have added a section on personnel protection because this is our highest priority in the design and operation of power systems. Thus, our topics are the protection of people, power converters, and magnet loads (protected from the powering equipment), including normal conducting magnets and superconducting magnets.

## 1 General protection techniques

In the protection of any of these topics, there are a number of general techniques that we use to help us achieve our protection goals. We will list them below and then point out when they are used in a variety of situations.

### 1.1 Redundancy

This means designing two independent paths leading to the desired protective action. For example, the overcurrent protection circuitry of a rectifier power converter might consist of a current transductor monitor circuit that turns off the thyristors in the power converter, as well as a shunt current monitor that opens the power-converter circuit breaker when the maximum current is exceeded.

A properly designed redundant protection system should have no elements that are common to both protective paths, since a failure of the common element might override both sources of protection.

### 1.2 Fail-safe design

This means that in the design of the protective circuitry, one should make sure that circuits are designed in such a way that any circuit failures that can be anticipated cause the system to turn OFF. The most common example of this is in the thermal protection of loads. We typically use a thermal switch that becomes an open circuit when the temperature is exceeded. The switch is connected to circuitry that turns off the power system when an open circuit is sensed. This avoids the anticipated problem of someone leaving the thermal switch disconnected from the system during installation or maintenance periods. If this does occur, the system cannot be turned on.

### 1.3 Response to power outages

As it is a virtual certainty that the a.c. power to the system will drop out during operation, it is important to design the electronics to respond in a safe and protective way to this situation. A common way to deal with this is to run the control system from the a.c. source of an uninterruptible power supply. This battery-backed system maintains control voltages intact until power elements can be safely de-activated.

### 1.4 Testing protection circuits

Well-designed protective circuits cannot be trusted until they are tested in place. This will involve opening doors to see that door-interlocks work, disconnecting or heating thermal switches, tripping the breaker supplying a.c. power to the system, etc. Redundant systems must be tested individually by bypassing each of the paths to see that the other path works.

### 1.5 Trouble-shooting aids

Usually when a power converter or magnet system trips off, this happens because an interlock has detected an improper situation and has acted correctly. Typically, it does not mean that the system is broken and in need of repair.

It is important to understand these trips and deal with potential problems before they lead to serious failures or endless annoyances.

To this end, it is important to ensure that all trip indications latch and identify themselves. Beyond this, modern day transient recorders that can record the sequence of events leading to a trip are an invaluable tool, especially in complex systems.

### 1.6 Self-contained protection

This means that the power system's internal controls must prevent incorrect commands from the accelerator control system from putting the power system in an unprotected state. For example, an operating current reference beyond the specified maximum current should be rejected or clamped.

## 2 Protection of people

### 2.1 Lock and tag out system

Our primary method of protecting people who will be working on a power system is our lock and tag out (LOTO) system. Before people are allowed to touch any element of the power system, they must first follow the LOTO procedure to turn the system off, lock out the sources of power, verify the absence of input power, discharge the energy stored in capacitor banks, and install ground clamps where necessary to ensure that nothing can become charged after the LOTO is completed. Installed locks are tagged with the name of the person who installed them and can only be removed by that person.

### 2.2 Interlocks

We use interlocks on power cabinet doors and in the accelerator tunnel to make sure that if someone unwittingly opens the door to a hazardous area, the power system will trip off. The main hazards will be removed from the equipment, but since the full LOTO has not been accomplished, the equipment will not yet be approved for access.

### 2.3 Captured key systems

In equipment with unusually high hazards, we often use a captured key system, which requires a person to lock off the source of power to the cabinet before gaining access to the key that will allow access to the cabinet. This does not take the place of the LOTO procedure; the LOTO procedure must be followed before one can work inside the system cabinet.

# 3 Protection of power converters

The most common types of power converter used in magnet systems are the rectifier power converter and the switch-mode power converter. In this paper, we will focus on protecting the rectifier power converter.

## 3.1 Overcurrent protection in a.c. systems

As in a small bench supply, a high-power rectifier power converter must be protected from the overcurrent that will occur following an internal short circuit. This protection is typically provided by a circuit breaker capable of interrupting the high fault current and disconnecting the power converter from the power line (see Fig. 1).

There are two levels of fault current that must be handled. If the fault occurs on the secondary of the rectifier transformer, the fault current on the primary is limited by the 'impedance' of the rectifier transformer (essentially its leakage inductance) and is typically 20 times the maximum operating current.

If the fault occurs on the primary side of the rectifier transformer, the current is limited by the 'impedance' of the transformer (typically at the power substation) powering the feeder system to which the power converter is attached. This fault current is usually much higher than the first case. The power-converter circuit breaker must be specified to interrupt this fault current in a system that is properly 'coordinated'.

## 3.2 Overvoltage protection in a.c. systems

The rectifier transformer is typically protected by 'surge suppressors' located from the primary windings to ground, which limit the voltage if the feeder system is hit by lightning or if there is some other source of large transient voltages.

The transformer secondary is protected from transients that occur when the breaker opens under load by the use of RC (resistor–capacitor) snubber networks.

The transformer and circuit breaker are specified and tested according to industrial standards to be able to withstand high transient voltages (e.g., 110 kV impulse testing on 15 kV rated equipment).

## 3.3 Protection in d.c. systems

The most important elements to protect are the thyristors (silicon-controlled rectifiers). These devices must be chosen to have ample voltage ($2.5 \times$ operating voltage) and current ratings. Using the thyristor data sheets, we calculate the maximum temperature the thyristor junction will attain during the most severe operating mode. We select a device whose junction temperature we can keep below 80°C.

We also calculate the temperature variation of the junction in ramping systems and make sure that the temperature cycling is less than 30°C in systems whose cycle time is greater than 1 s. These two limits, which relate to the power-converter current, ensure a long lifetime for the thyristors.

The d.c. side of the power converter often has a passive filter to reduce ripple voltage on the rectifier output. We protect the filter choke, which is usually water-cooled in high current systems, with a thermal switch. The filter capacitors often have overpressure switches, which detect a failure and are used to interlock the power converter.

## 3.4 Response to loss of power

We usually have an uninterruptible power supply to maintain the control voltages in the event of a power failure. This allows us time to bypass the current in the supply into the bypass thyristor (SCR) and then open the circuit breaker.

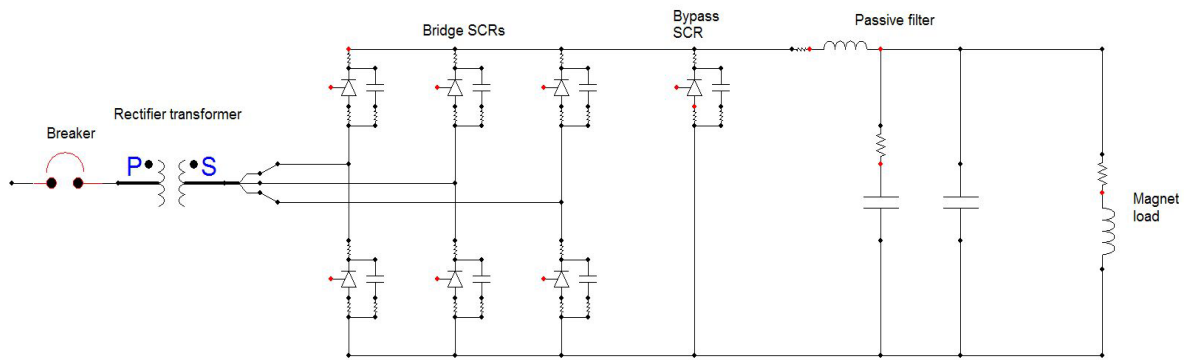

**Fig. 1:** Rectifier power-converter circuit

## 4 Protection of conventional magnets

### 4.1 Overcurrent protection

We normally use a direct current–current transformer (DCCT) to measure and regulate the load current coming out of the power converter. The same DCCT signal is compared with a trip threshold and bypasses the converter when the level is exceeded.

This protection is backed up with a shunt measurement that causes the circuit breaker to trip when the measured current exceeds the threshold, which is usually set slightly higher than the DCCT threshold.

For non-d.c. loads, it is sometimes necessary to design a circuit that trips the power converter based on the root mean square value of the current.

### 4.2 Voltage-to-ground protection

Magnets are manufactured with electrical conductors wound around an iron core, and isolated from the grounded core by a system of insulating material. The insulation system is designed to withstand up to a minimum voltage between the conductors and the core without breaking down.

The design and validation of the magnet insulation system must be coordinated with the worst-case voltages that magnets will experience while operating in their circuit.

The 'ground fault circuit' in a power converter has two functions: to detect unwanted current going to ground and to minimize the voltage-to-ground of the magnet load. We will consider the simple case of one power converter and one magnet as its load.

Figure 2 shows this simple case with two common forms of ground fault detector: the fused detector and the balanced high-impedance detector. In the former kind, the negative terminal of the power converter is held at ground with a low current fuse. If a short circuit to ground develops towards the positive terminal of the converter, it wants to pull the negative terminal below ground, but the fuse conducts to maintain the terminal at ground potential until there is sufficient current to blow the fuse. A circuit then detects that the fuse is blown and turns off the power converter.

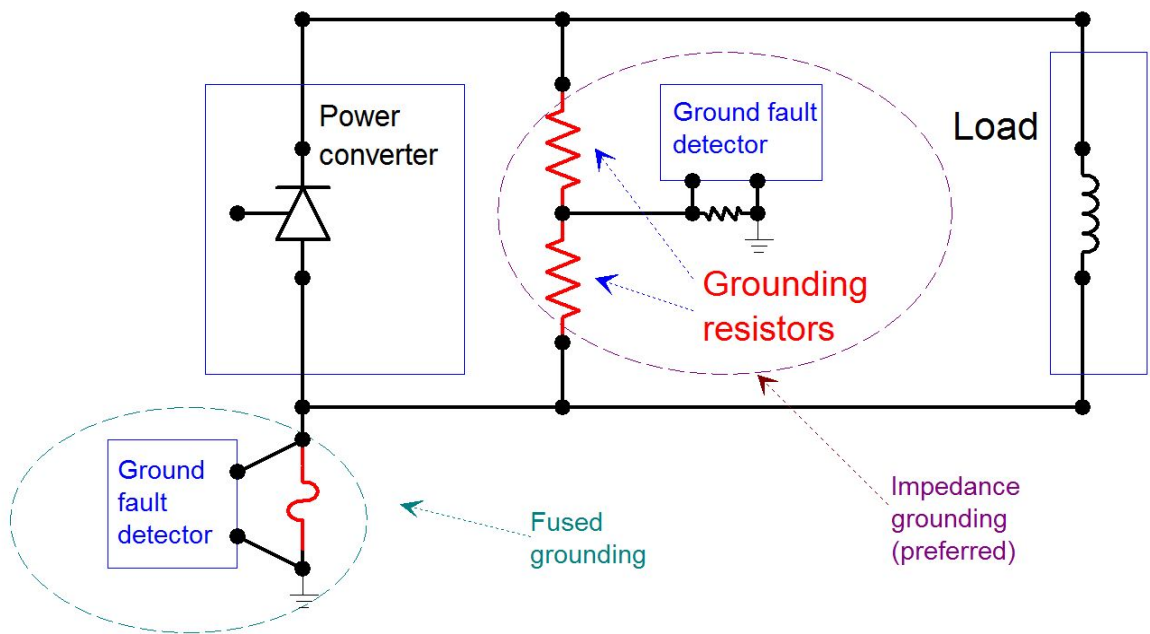

**Fig. 2:** Grounding schemes

This circuit has two drawbacks. If the power converter is a 1 kV unit, the part of the load connected to the positive terminal always sees the thousand volts to ground, stressing the magnet insulation at that level. Also, when a ground fault does develop, a high current may want to flow into the fuse and may exceed its interrupting rating. If this happens, this large current will flow through the faulted point for a long time and damage the point beyond recovery.

The balanced high-impedance grounding detector in Fig. 3 has grounding resistors of the order of a few kilo-ohms, so it cannot conduct high ground currents. This circuit also balances the voltage-to-ground, so that the maximum voltage that any part of the magnet sees in normal operation is ±500 V. If a ground fault develops in this system, the summing point between the two grounding resistors, which normally stays close to ground, is in deflected either a positive or negative direction, and this change in its voltage causes a trip of the power converter.

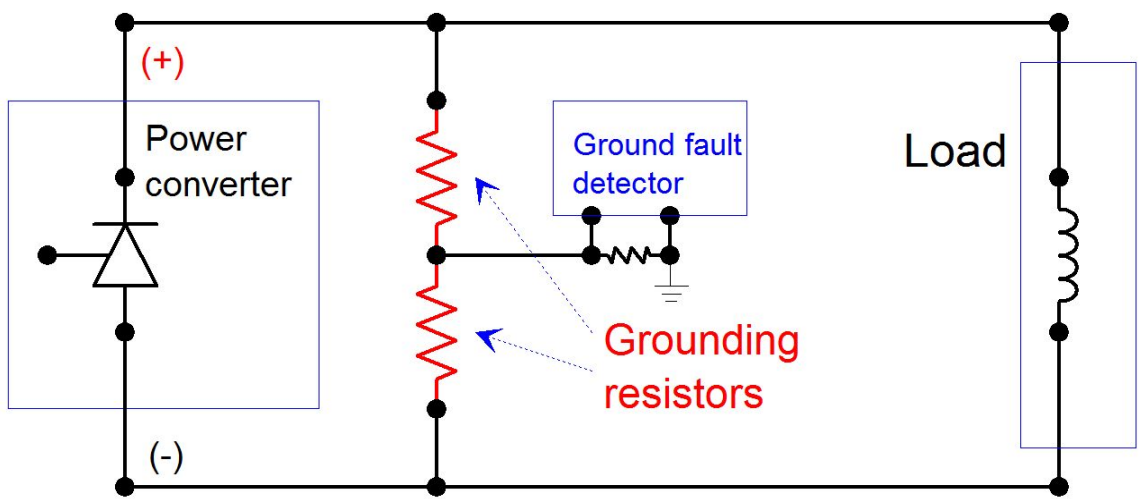

**Fig. 3:** Balanced high-impedance grounding

In this preferred scheme, although the maximum operating voltage-to-ground is half of the power-converter voltage, if a ground fault occurs near the negative terminal of the power converter, its positive terminal will be raised to the full voltage of the power converter before the ground fault circuit can react. If a second ground fault occurs there, this is called a 'double ground fault'. This is very much to be avoided because it allows large, destructive currents to go through both ground points. Thus, the magnet insulation in this example must be designed to withstand more than the full power-converter voltage (with some margin), so that a double ground fault is highly unlikely.

## 5  Protection of superconducting magnets

Table 1 provides details for some example superconducting magnets. Circuits with superconducting magnets must have the same overcurrent and voltage-to-ground protection systems that are required for conventional magnet circuits. In addition, these circuits must have a quench protection system to protect the magnets when they lose their superconducting properties. We will start with some of the definitions and features of superconducting magnets.

### 5.1  Superconducting cables and magnets

– Magnets are most commonly wound with superconducting cable (niobium–titanium).

– Superconducting cable becomes superconducting (zero resistance) below a critical temperature (usually <10 K). Low temperatures are established and maintained in a bath of liquid helium that is cooled in a cryogenic refrigeration system.

– Superconducting cable is made of several wire strands, each made of many superconducting filaments within a copper matrix.

– The cable has a 'short sample' maximum current, beyond which it will 'quench' and lose its superconducting properties.

– The copper matrix stabilizes the cable and provides an alternate current path for a short time when the superconductor 'quenches' or leaves its superconducting state.

– A magnet wound with the wire has a lower maximum current because magnetic fields within the magnet decrease the cable's maximum conduction.

– A magnet's maximum current can be increased by reducing its temperature. (Example: Fermilab's Tevatron magnets went from 4000 A to 4400 A when the helium temperature was reduced by 0.75 °K at a cost of $6 million).

**Table 1:** Different accelerator magnets with their respective operating limits and the cross-sectional areas of their cables (the MIIT limit column is explained later in the text).

| Magnet | Short sample limit, kA | Maximum operating current, kA | Operating temperature, K | MIIT limit, $A^2 s$ | Cable cross-section, $mm^2$ |
|---|---|---|---|---|---|
| LHC: main bend | 13 | 11.5 | 1.85 | $32 \times 10^6$ | 22 |
| LHC: main quadrupoles | 13 | 12.1 | 1.85 | $32 \times 10^6$ | 22 |
| LHC: 600 A | 0.6 | 0.55 | 1.85 | $50 \times 10^3$ | 1 |
| Tevatron: dipoles | | 4.4 | 4.5 | $7 \times 10^6$ | 10 |

# 6 Quenching

A magnet conducting current in superconducting mode at cryogenic temperatures can suddenly lose its superconductive state, usually beginning at a particular spot in the magnet cable, when something causes the temperature at that spot to rise above the critical temperature.

Once the initiating spot quenches, the heat generated from the resistance typically keeps it in the quenched state, and the quenched area spreads to nearby areas at a speed known as the 'quench velocity'.

## 6.1 Causes of quenching

There are several well-known causes of quenching in accelerator magnet systems.

### 6.1.1 Training

Training refers to the small slippage of one of the superconducting cables within the tightly clamped magnet structure. This slippage occurs when the magnet undergoes its initial powering. The slippage generates some heat, by friction, and the heat can cause a quench at that location. Usually, the cable slips into a more stable position and, once the magnet has undergone a few quenches as the current is increased to its maximum operating value, these quenches no longer occur; we say that the magnet has been 'trained'.

### 6.1.2 Excess d$I$/d$t$—eddy currents

Fast current changes can induce local eddy currents within the superconductor cable itself. These can cause local heating, leading to a quench of the cable. Sometimes, the action of rapidly decreasing the current in a magnet system to protect an element in the circuit can cause other elements to quench from excess d$I$/d$t$.

### 6.1.3 Particle beam heating

If the particle beam in a storage ring becomes unstable or the injection or extraction of the beam are not well controlled, part of the beam can hit the superconducting cable and cause a quench.

### 6.1.4 Cooling system problems

If the helium warms to above the critical temperature, quenches will occur.

### 6.1.5 Exceeding the short sample limit

Normally, the powering system is set up to protect against this happening, but sometimes the protection does not work correctly.

### 6.1.6 Spontaneous quenches—unknown origins

Mystery quenches do happen in complex systems. This is why it is important to have good monitoring systems to provide the best chance of understanding each quench.

## 6.2 Heating of the initiating spot—MIITs

The initiating spot starts to heat first, and is normally the hottest place in the quenching magnet. Keeping its ultimate temperature below a damaging level (450 K) is critical in protecting the quenched magnet.

A simplified way of thinking about the temperature rise at the initiating spot is to imagine it as a length of copper wire weighting $M$ (g) with resistance $R$ (Ω) and specific heat $C$ (J/(g K)). Then the adiabatic temperature rise of the spot will be:

$$\Delta T = R \frac{\int I^2 \, dt}{MC} = \frac{R}{MC} \int I^2 \, dt \ .$$

Note that the temperature rise in this calculation is independent of the wire length.

You can calculate an integral of $I^2$ that will raise the temperature of the initiating spot from 10 K to 450 K. This integral is called the 'MIIT' limit of the cable. Usually this integral is in 'millions of amp squared seconds', hence the term 'MIIT'.

This simplified example assumes constant values of $R$ and $C$. In real life, the resistance of copper varies by about a factor of 100 between cryogenic and room temperatures and the specific heat varies by about a factor of 300. The calculation of realistic MIITs is not so different from the ideal because both $R$ and $C$ increase with increasing temperatures, and thus tend to compensate each other.

The MIITs calculation is crucial for determining the two primary factors of the quench protection system:

– How quickly the current in the magnet must be reduced once the quench is detected;
– How quickly a quenched must be detected once the initiating spot quenches.

The allowable MIIT (see Table 2) in a magnet and the maximum operating current in its circuit set a maximum time-scale for the reduction of the current once a quench has been detected. Some examples of this relationship follow.

**Table 2:** MIIT limits and operating currents for three accelerator circuits: kIIT, 'thousands of amp squared seconds'; MIIT, 'millions of amp squared seconds'.

| Accelerator circuit | MIIT limit | Operating current, kA |
|---|---|---|
| Tevatron dipole | 7 MIIT | 4 |
| LHC dipole | 32 MIIT | 12 |
| Tevatron correction coil | 3.2 kIIT | 0.05 |

The time-scales for the reduction of the currents in these circuits are:

– For a Tevatron dipole running at 4 kA (16 MIIT/s), the current in the quenching magnet must be substantially reduced within 0.44 s (7 MIIT ÷ 16 MIIT/s).
– For an LHC dipole running at 12 kA (144 MIIT/s), the current in the quenching magnet must be substantially reduced within 0.22 s (32 ÷ 144).
– For a Tevatron correction element running at 50 A (2.5 kIIT/s), the current in the quenching magnet must be substantially reduced within 1.28 s (3.2 ÷ 2.5).

### 6.2.1 *Methods to reduce number of MIITs following quench detection in magnet circuits*

– Reduce power-converter voltage to zero if cable resistance is enough to limit the MIITs (example: Tevatron extraction quadrupole loops).
– Reduce power-converter voltage and use energy extraction circuit to insert a resistance sufficient to limit the MIITs (example: Tevatron main quadrupole correction loop).
– Reduce power-converter voltage and fire 'Heaters' (example: Tevatron low-beta magnets).

### 6.2.2 *Methods for limiting MIITs after quench detection*

#### 6.2.2.1 *Method a*

Reduce power-converter voltage to zero if cable resistance is sufficient to limit the number of MIITs (e.g., Tevatron extraction quadrupole loops). See Fig. 4.

System data:
$$\text{Four magnets at } 0.46 \text{ H each} = 1.84 \text{ H},$$
$$\text{Cable } R = 2.45 \text{ Ω},$$
$$L/R = 0.75 \text{ s},$$
$$I_{\max} = 50 \text{ A},$$

For exponential decay,
$$\text{IIT} = \frac{1}{2}\left(I_{\text{pk}}^2 \tau\right),$$
$$\text{IIT} = \tfrac{1}{2} \times 50 \times 50 \times 0.75 \text{ s} = 0.94 \text{ kIIT},$$

which compares well with the 3.2 kIIT limit.

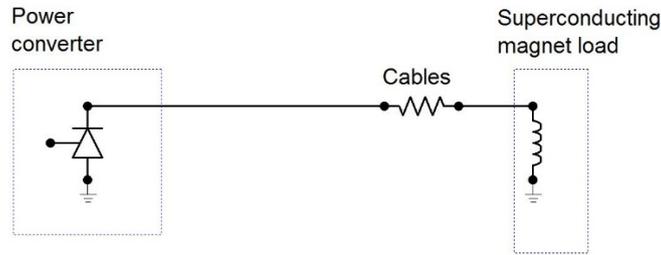

**Fig. 4:** Limiting MIITs after quench detection, method a

### 6.2.2.2 Method b

Reduce voltage and use energy extraction circuit (insert a resistance, as in the Tevatron main quadrupole correction coil loops). See Fig. 5.

System data:
$$90 \text{ magnets at } 0.46 \text{ H each} = 41 \text{ H},$$
$$\text{Cable } R = 5.8 \text{ Ω},$$
$$\text{Cable } L/R = 7.14 \text{ s},$$
$$\text{Dump } R = 20 \text{ Ω},$$
$$L/R = 1.6 \text{ s, including dump } R,$$
$$I_{\max} = 50 \text{ A}$$
$$\text{IIT} = \tfrac{1}{2} \times 50 \times 50 \times 7.14 \text{ s} = 8.9 \text{ kIIT without dump } R,$$
$$\text{IIT} = \tfrac{1}{2} \times 50 \times 50 \times 1.6 \text{ s} = 2.0 \text{ kIIT with dump } R.$$

The result with the dump resistance compares well with the 3.2 kIIT limit.

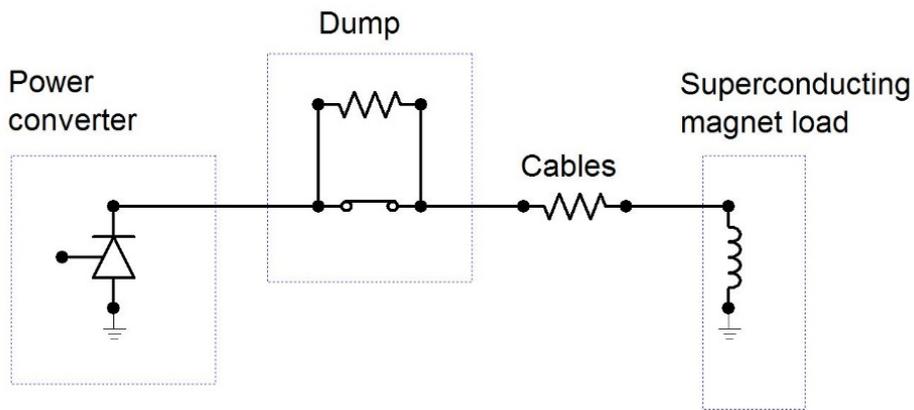

**Fig. 5:** Limiting MIITs after quench detection, method b

*6.2.2.3 Method c*

Reduce voltage and fire heaters (e.g., Tevatron low-beta magnets). See Fig. 6.

System data:

$$\text{Two magnets at 46 mH each} = 92 \text{ mH},$$

$$\text{Cable } R = 3 \text{ m}\Omega;$$

$$\text{Cable } L/R = 30 \text{ s},$$

$$\text{Magnet } R = 375 \text{ m}\Omega \text{ equivalent with heater firing},$$

$$\text{Magnet, with heater firing, } L/R = 0.25 \text{ s},$$

$$I_{\max} = 5 \text{ kA},$$

$$\text{IIT} = \tfrac{1}{2} \times 5000 \times 5000 \times 30 = 375 \text{ E6} = 375 \text{ MIIT}$$

With heater firing,
$$\text{IIT} = \tfrac{1}{2} \times 5000 \times 5000 \times 0.25 = 3.13 \text{ E6} = 3.13 \text{ MIIT}$$

This result compares well with the 3.2 MIIT limit.

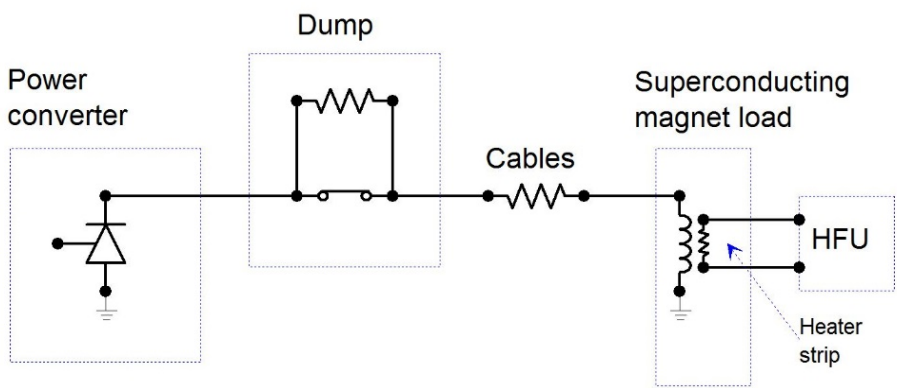

**Fig. 6:** Limiting MIITs after quench detection, method c: HFU, heater firing unit

Heaters are steel strips that are pressed against the outer windings of a magnet. They are designed to have a reasonably low thermal insulation but sufficient electrical insulation to avoid arcing to the magnet winding. When a quench is detected, the heater firing unit is triggered and discharges the energy in a capacitor bank into the steel strip. This energy is sufficient to initiate a quench in a large fraction of the magnet cabling. The increase in resistance of this large volume of quenched cabling is sufficient to reduce the magnet current before it reaches its MIIT limit. Details of two examples of heater circuit are given in Table 3.

**Table 3:** Two examples of heater circuit

| Circuit | Capacitance, mF | Voltage, V | Energy, kJ | Strip resistance, $\Omega$ | Discharge time, ms |
|---|---|---|---|---|---|
| Tevatron dipole | 6.6 | 450 | 0.67 | 20 | 18 |
| LHC dipole | 7 | 900 | 2.8 | 12 | 84 |

Although heaters are an effective way to induce a large-scale quench in a superconducting magnet, the technique requires extra strips to be built into the magnet and the use of a high-energy discharge circuit. Most concerning in this approach is the possibility of heaters failing and, in the worst case, shorting the magnet winding to ground. Although heaters have certainly failed, we are not aware of cases in which the magnet was shorted.

A new induced-quenching technique is being developed at CERN by E. Ravaioli. It is called coupled loss-induced quenching, and it involves discharging a capacitor bank across the terminals of the magnet, causing a high-frequency ringing and inducing a d$I$/d$t$ quench.

### 6.2.3 *The special case of series magnet strings with large quantities of stored energy*

The accelerator world often contains extended systems with many magnets and large quantities of stored energy.

**Table 4:** Two examples of extended multi-magnet systems

| Accelerator | Number of magnets | Maximum current, kA | Total inductance, H | Energy, MJ |
|---|---|---|---|---|
| Tevatron ring | 776 | 4.4 | 30 | 290 |
| LHC dipole sector | 154 | 11.5 | 15.4 | 1018 |

It is impractical to remove this much energy from the magnet systems in a fraction of a second, so an approach using heater firing, 'bypassing', and 'energy extraction' has been used. When a quench is detected in one of the magnets, the quench protection system takes the three actions:

– fire the heater firing unit on the quenching magnet;

– establish a bypass path for the main circuit current to go around the quenching magnet while its own current decays within a fraction of a second;

– open switches to insert dump (energy extraction) resistors so that the magnet circuit current can decay on a time-scale of several seconds.

The time constant of the dump is coordinated with the number of MIITs that the bypass path can absorb without overheating. The time constants used in the two cases listed in Table 4 are:

$$\text{Tevatron} = 12 \text{ s},$$

$$\text{LHC dipole sector} = 100 \text{ s}.$$

Fig. 7 is a composite showing the quench protection systems of the LHC and the Tevatron. Both systems use heater firing units, as described before.

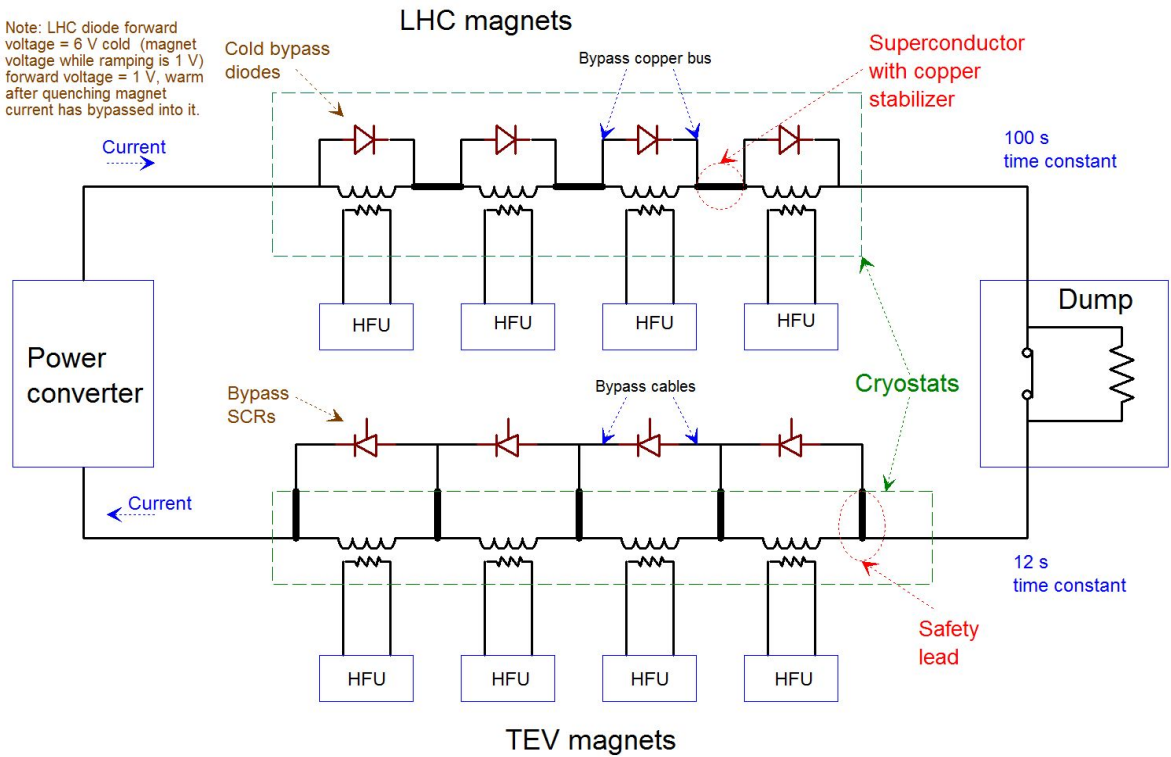

**Fig. 7:** Tevatron and LHC quench protection systems

In the Tevatron, the bypass circuit goes through a silicon-controlled rectifier (thyristor) that is triggered only when a quench is detected. The current in the quenching magnet exits through copper 'safety leads', is carried to the external silicon-controlled rectifier bypass switch via copper cables, and returns on the other side of the quenching magnet in the same way. The limiting elements of the bypass circuit are the safety leads, which are made as small as possible to limit heat leaks from the cryogenic system to the outside. These leads can handle the peak magnet current (4.4 kA) during a 12 s exponential decay without overheating.

Figure 8 shows the redundant features of the Tevatron bypass switch system. It features parallel silicon-controlled rectifiers, parallel firing circuits, and parallel uninterruptible power supply systems.

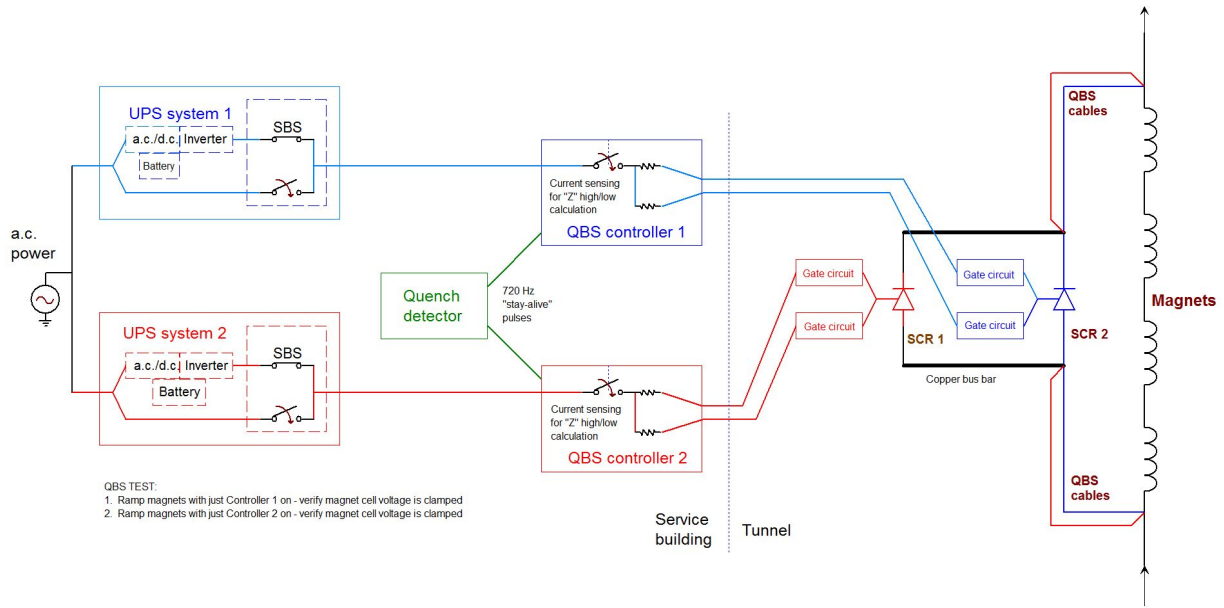

**Fig. 8:** Tevatron quench bypass system, emphasizing redundant architecture. QBS, quench bypass switch; SBS, standby power supply; SCR, silicon-controlled rectifier; UPS, uninterruptible power supply.

Figure 9 shows what happens to the current in a quenching Tevatron magnet in a string with a bypass silicon-controlled rectifier. We have measured the decay time of the magnet current as less than 0.2 s. This indicates that the resistive drop of the heater-driven magnet builds up to 650 V. The current is driven out of the magnet with an amp squared seconds integral of 5 MIITs.

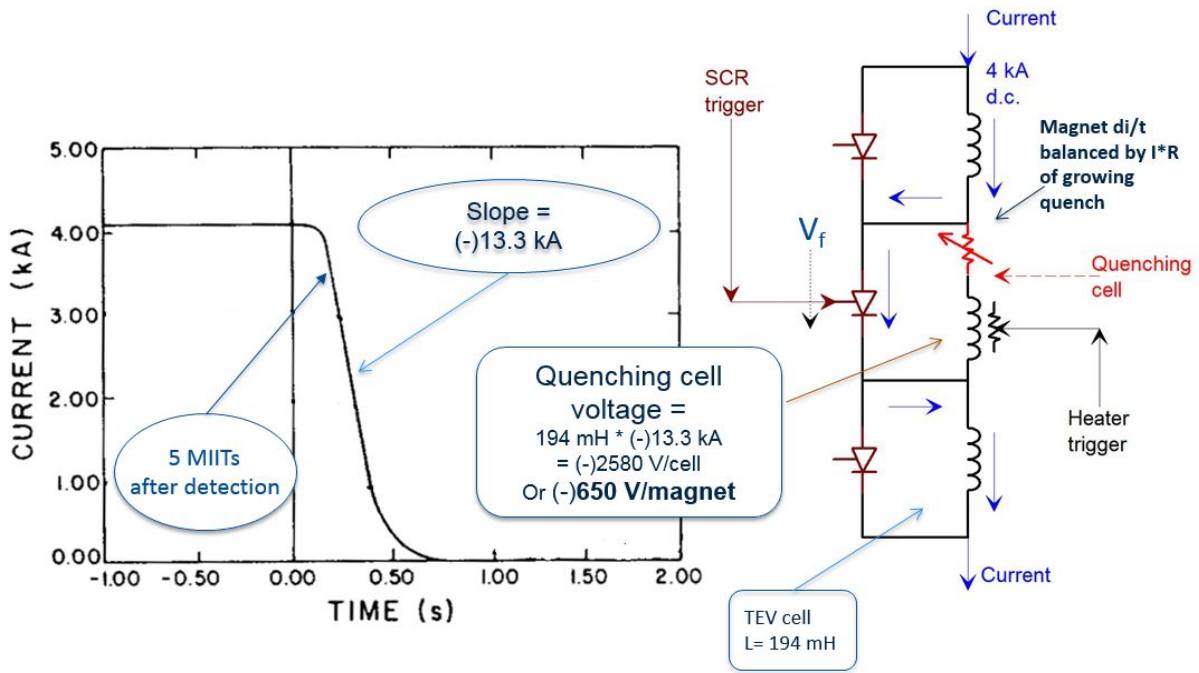

**Fig. 9:** What happens to the current in a quenching Tevatron magnet in a string with a bypass silicon-controlled rectifier when the heaters are fired and the current is bypassed.

In the LHC, the bypass circuit goes through the 'cold bypass diode', which starts conducting when the voltage across the quenching magnet exceeds 6 V. At this point, the diode's self-heating rapidly transforms it into a normal diode that conducts at 1 V. It then bypasses the quenched magnet current through a copper bus. This 'copper stabilizer bus' is a 1.5 cm× cm copper bus with a slot inside it that carries the superconductor cable between magnets. This stabilizer bus must carry the bypass current if a section of interconnecting bus quenches. The limiting elements of the bypass circuit are the copper stabilizer buses. They can handle the peak magnet current (12 kA) during a 100 s exponential decay without overheating.

Table 5 gives a summary of MIITs that are deposited in various systems once the energy extraction circuits have been triggered.

**Table 5:** MIITs that are deposited in various systems once the energy extraction circuits have been triggered

|                                         | MIITs      | Maximum   |
|-----------------------------------------|------------|-----------|
| Tevatron extraction quadrupole          | 0.9 kIIT   | 3.2 kIIT  |
| Tevatron main quadrupole correction loop| 2.0 kIIT   | 3.2 kIIT  |
| Tevatron low-beta magnets               | 3.2 MIIT   | 5.2 MIIT  |
| Tevatron dipoles                        | 5.0 MIIT   | 7.0 MIIT  |

# 7 Quench detection

To trigger the energy extraction circuits to limit the MIITs in a magnet, we must detect that a quench is occurring in that magnet.

What are we detecting? Basically, the extra *IR* 'resistive' voltage that should not be there in a superconducting load ($R$ = resistance of quenching cable as quench propagates).

How much time do we have? The time between the initiation and detection of a quench, $T$.

– Remember, MIITs start accruing from moment that the initiating spot quenches.
– Once the quench is detected and the protection system responds, a certain number of (after energy extraction) MIITs will be deposited.
– Therefore, the maximum time Tmax is given as:

$$T_{\max} = \frac{(\text{Maximum MIIT – MIIT after energy extraction})}{I^2}.$$

What voltage must we detect in the allowed time? This depends on the quench velocity within the magnet. We need to make calculations and make measurements to determine what sensitivity our quench detection system must have. Two examples will illustrate the process. Table 6 provides details for some quench detection sensitivities.

## 7.1 Example A: The quadrupole correction loop

Max = 3.2 kIITs; after energy extraction = 2.0 kIITs,

$$\Delta T = \frac{\text{IITs(rating)} - \text{IITs(after energy extraction)}}{I(\text{before quench})^2},$$

Time to detect, $\Delta T = \frac{3.2 \text{ kIITs} - 2.0 \text{ KITTs}}{50^2} = 0.48 \text{ s}.$

### 7.1.1 *How much voltage? How quickly does resistive voltage increase?*

Experiments were performed on dipole correction elements during which a heater was fired to cause a quench condition while 50 A was being conducted through the element. The coil voltage reached 10 V within approximately 0.25 s. This compares well with the calculated detection time 0.48 seconds.

Therefore, a 10 V detection threshold would be sufficient. We were able to operate with a threshold of 4 V without nuisance trips.

## 7.2 Example B: The main Tevatron loop

After energy extraction, 5 MIIT are deposited when running at 4 kA.

### 7.2.1 How much time do we have?

$$\text{Time to detect, } \Delta T = \frac{(7\,\text{MIITs} - 5\,\text{MIITs})}{4^2} = 0.125 \text{ s}.$$

### 7.2.2 How quickly does resistive voltage increase?

A complicated calculation involving quench velocity in single wires, reveals that we must choose a threshold of 0.5 V. In practice, we aim for a stringent goal and see how close we can come, within the constraints of periodic and random noise. We must have a large enough margin to avoid nuisance trips.

**Table 6:** Summary of various quench detection sensitivities

| System | Trip threshold, V | Averaging time, ms |
|---|---|---|
| Tevatron quadrupole correctors | 4.0 | 10 |
| Tevatron main dipoles | 0.5 | 50 |
| LHC main dipoles | 0.1 | 10 |
| LHC 600 A circuits | 0.4 | 200 |

## 8 How do you detect resistive voltage?

### 8.1 Compare voltage across similar magnets in series

#### 8.1.1 Example A: Tevatron correction quadrupole loop

Compare the voltage across one set of 45 magnets with that across the other set of 45 magnets (carefully, using centre tap). Look for a 4 V difference.

#### 8.1.2 Example B: Tevatron main loop

Compare voltage across four cells (five magnets each). Look for a 0.5 V difference.

#### 8.1.3 Example C: LHC 13 kA dipole circuits

This is illustrated in Fig. 10. Look for a 0.1 V difference.

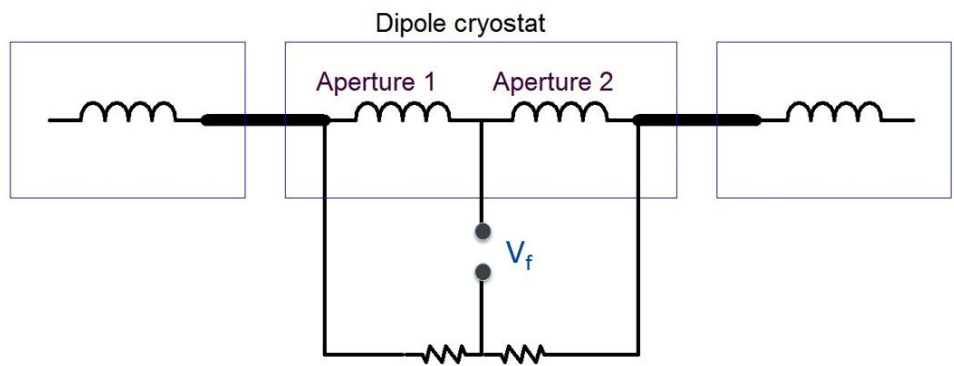

**Fig. 10:** Detecting resistive voltage in an LHC 13 kA dipole circuit

## 8.2 Compare $L \times dI/dt$ or magnet with measured magnet voltage

### 8.2.1 Example: LHC 600 A corrector circuits

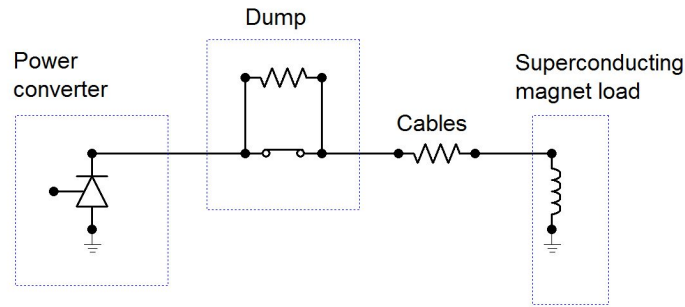

**Fig. 11:** Comparing $L \times dI/dt$ or magnet with measured magnet voltage for an LHC 600 A corrector circuit

This is illustrated in Fig. 11. Subtract:

$$V_{\text{mag}} - L\frac{dI}{dt}.$$

Look for a 0.4 V difference.

Find the quench voltage, $V_q$:

$$V_q = V_{\text{mag}} - Lm\frac{dI}{dt}.$$

This method is sensitive to noise on the measured current signal and to the complex impedance of the magnet. (The LHC is still upgrading these quench detection systems.)

## 8.3 Comparing quench detection method

The method of comparing similar magnet voltages with each other typically allows for lower quench detection thresholds:

- no $dI/dt$ noise issues;
- no complex magnet impedance issues.

Comparing only two magnet voltages with each other introduces a vulnerability to symmetrical quench growth in both magnets. The LHC encountered this in the main dipole bus and mitigated the possibility by installing an additional system, which compared four magnets with each other.

# 9 Unexpected problems

## 9.1 Large Hadron Collider

Bad solder joint in the copper stabilizer used in the main dipole bus splices (Fig. 12) could not support the current when the superconducting cable in the splice quenched. The resulting arc deposited huge amounts of energy into the cryogenic system, causing excessive pressure, and leading to a 'helium leak'.

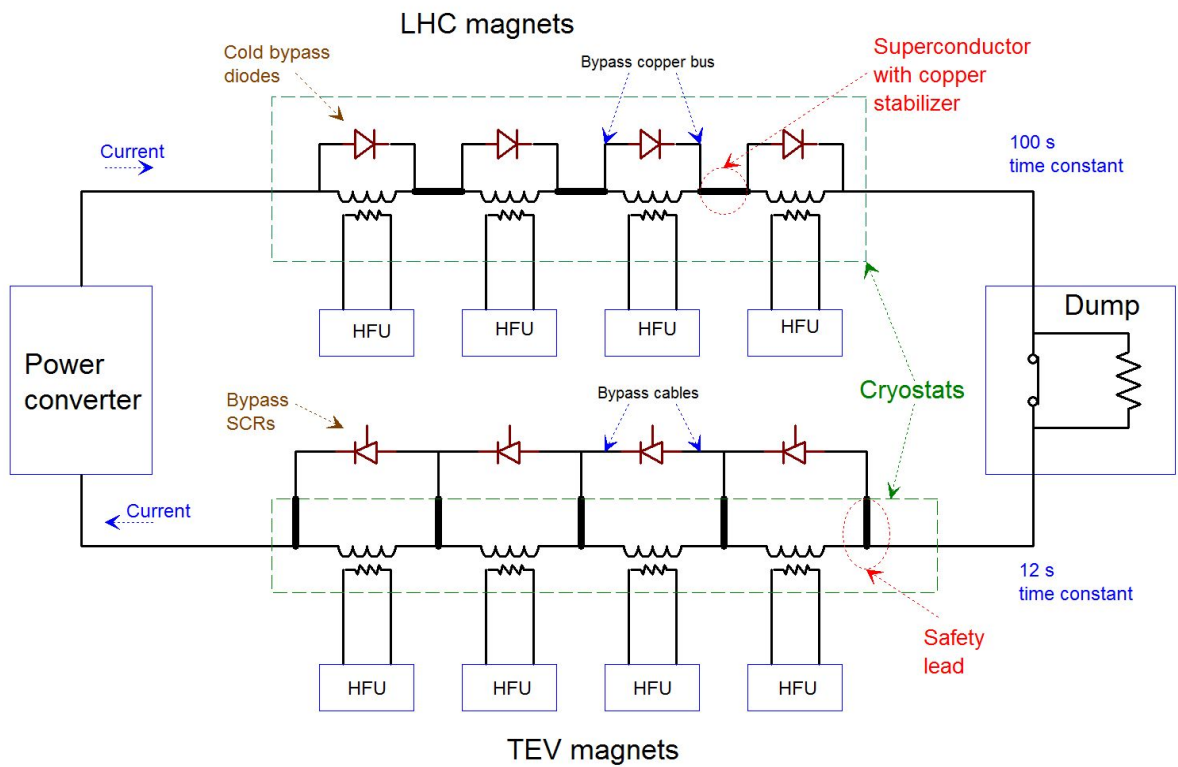

**Fig. 12:** Arrangement of LHC and Tevatron magnets: HFU, heater firing unit; SCR, silicon-controlled rectifier

## Definitions

**Cable short sample curve**: The two-dimensional curve in magnetic field and current space formed by the intersection of the critical surface and a plane of constant operating temperature, where the current density is integrated over the cross-section of a specific cable. This curve is measured with a 'short sample' of the cable placed in different magnetic fields while the current is slowly increased, until a quench occurs.

**Critical surface, cable short sample curve, magnet short sample limit**: These terms, which are directly related but distinctly different, are often referred to by slightly different or abbreviated names. Starting with the most general term, 'critical surface', each term is increasingly more specific and less general.

**Critical surface**: The three-dimensional surface in temperature, magnetic field, and current density space under which a specific conductor remains superconducting. The points where this surface intersects the three axes are called the critical points; $T_c$, $B_c$, and $J_c$, respectively.

**DCCT**: Direct current–current transformer

**Dumping**: Process of inserting resistors into a circuit consisting of superconducting elements to remove stored energy

**HFU**: Heater firing unit

**Lower critical field**: The magnetic field at which the magnetic flux starts to penetrate a type 2 superconductor

**Magnet short sample limit**: The current where the magnet (peak field) load line intersects the cable short sample curve

**Magnet short sample margin**: The difference between the operating current and the magnet short sample limit

**Magnet temperature margin**: The temperature elevation necessary to diminish the magnet short sample margin to zero

**MIIT**: The exact (adiabatic) relationship between MIITs and temperature depends on only two things, the intrinsic conductor material properties and the cross-sectional area squared:

$$A^2 D \int_{T_0}^{T} \frac{C(T)}{\rho(T)} dT = \int_{0}^{\infty} I(t)^2 \, dt = \text{MIIT} \ .$$

**MOV**: Metal oxide varistor, a non-linear device for controlling overvoltages

**Power converter**: Any device that converts one form of voltage or current to another form. In this context, usually a power supply that converts an incoming a.c. line to d.c.

**QBS**: Quench bypass switch

**Quench**: Sudden runaway loss of superconductivity, driven by the heat of normal conduction, driven by the loss of superconductivity, driven by the heat of normal conduction, etc.

**SCR**: Silicon-controlled rectifier, a solid state switch, in which applying a voltage to the 'gate' will switch the device from an open circuit to a diode

**Superconductivity**: A phenomenon of exactly zero electrical resistance and expulsion of magnetic fields occurring in certain materials when cooled below a characteristic critical temperature

**Type 1 superconductors**: This category of superconductors mainly comprises metals and metalloids that show *some* conductivity at room temperature. They require incredible cold to slow down molecular vibrations sufficiently to facilitate unimpeded electron flow.

**Type 2 superconductors**: Except for elements vanadium, technetium, and niobium, the type 2 category of superconductors comprises metallic compounds and alloys. They achieve higher $T_c$ than type 1 superconductors by a mechanism that is still not completely understood. Current wisdom holds that it relates to the planar layering within the crystalline structure.

**Upper critical field**: The magnetic field (usually expressed in teslas (T)), which completely suppresses superconductivity in a type 2 superconductor at 0 K (absolute zero).